\documentclass[twocolumn]{aastex631}
\usepackage{float}

\usepackage{soul}

\newcommand{\msun}{M_{\odot}} 
\newcommand{\msunyr}{\msun\,{\rm yr}^{-1}} 

\newcommand{\mhe}{M_{\mathrm{He}}} 

\newcommand{\mco}{M_{\mathrm{CO}}}

\newcommand{\lsun}{L_{\odot}} 

\newcommand{\Teff}{T_{\mathrm{eff}}}

\newcommand{\rsun}{R_{\odot}} 

\newcommand{\mesa}{{\tt\string MESA}}
\newcommand{\gyre}{{\tt\string GYRE}}

\newcommand{\cp}{c_{\mathrm{p}}}

\newcommand{\tauth}{\tau_{\mathrm{th}}}

\newcommand{\tdyn}{\tau_{\mathrm{dyn}}}

\newcommand{\cs}{c_{\mathrm{s}}}

\newcommand{\alphaMLT}{\alpha_{\mathrm{MLT}}}

\newcommand{\numax}{\nu_{\mathrm{max}}}

\begin{document}

\title{The Asteroseismological Richness of RCB and dLHdC Stars}

\author[0000-0001-9195-7390]{Tin Long Sunny Wong}
\affiliation{Department of Physics, University of California, Santa Barbara, CA 93106, USA}

\author[0000-0001-8038-6836]{Lars Bildsten}
\affiliation{Department of Physics, University of California, Santa Barbara, CA 93106, USA}
\affiliation{Kavli Institute for Theoretical Physics, University of California, Santa Barbara, CA 93106, USA}

\correspondingauthor{Tin Long Sunny Wong}
\email{tinlongsunny@ucsb.edu}

\begin{abstract}

RCB stars are $L\approx10^4\,L_{\odot}$ solar-mass objects that can exhibit large periods of extinction from dust ejection episodes. Many exhibit semiregular pulsations in the range of $30-50$ days with semi-amplitudes of $0.05-0.3$ magnitude. Space-based photometry has discovered that solar-like oscillations are ubiquitous in hydrogen-dominated stars that have substantial outer convective envelopes, so we explore the hypothesis that the pulsations in RCB stars and the closely related dustless hydrogen-deficient carbon (dLHdC) stars, which have large convective outer envelopes of nearly pure helium, have a similar origin. Through stellar modeling and pulsation calculations, we find that the observed periods and amplitudes of these pulsations follows the well-measured phenomenology of their H-rich brethren. In particular, we show that the observed modes are likely of angular orders $l=0,1$ and $2$ and predominantly of an acoustic nature (i.e. $p$-modes with low radial order). The modes with largest amplitude are near the acoustic cut-off frequency appropriately rescaled to the helium-dominated envelope, and the observed amplitudes are consistent with that seen in high luminosity ($L>10^3\,L_{\odot}$) H-rich giants. We also find that for $T_{\mathrm{eff}}\gtrsim5400\,\mathrm{K}$, an HdC stellar model exhibits a radiative layer between two outer convective zones, creating a $g$-mode cavity that supports much longer period ($\approx 100$ days) oscillations. Our initial work was focused primarily on the adiabatic modes, but we expect that subsequent space-based observations of these targets (e.g. with TESS or Plato) are likely to lead to a larger set of detected frequencies that would allow for a deeper study of the interiors of these rare stars.

\end{abstract}


\section{Introduction} \label{sec:intro}

R Coronae Borealis (RCB) stars are a class of hydrogen-deficient carbon (HdC) stars \citep[for review, see, e.g.,][]{Clayton1996,Clayton2012}. 
While RCB stars normally have $M_{V}$ between $-3$ and $-5$ \citep[e.g.,][]{Alcock2001,Tisserand2009}, they often show large optical photometric declines (sometimes reaching 9 magnitudes) lasting years, caused by dust ejection episodes. 
Closely related to RCB stars are dustless hydrogen-deficient carbon (dLHdC) stars, which show RCB-like spectra but do not show dust ejection episodes \citep[e.g.,][]{Warner1967,Tisserand2022}. 
In this work, we do not distinguish between RCB and dLHdC stars, but use HdC as an umbrella term encasing both.

Two formation channels are proposed for RCB stars: the double-degenerate merger scenario \citep[][]{Webbink1984,Saio2002}, where an RCB star is born from the merger between a He white dwarf (WD) and a carbon-oxygen WD, and the final He flash scenario \citep[][]{Iben1996}, where a post-asymptotic giant branch star undergoes a late He shell flash. 
The former is favored in part because of the large ratio between $^{18}\mathrm{O}$ and $^{16}\mathrm{O}$ in RCB stars, which requires large amounts of $^{18}\mathrm{O}$ formed from partial He burning be dredged up to the surface \citep[][]{Clayton2007,GarciaHernandez2010}. 
In addition, recent 3D \citep[][post-merger evolution in 1D]{Munson2021} and 1D merger simulations \citep[][]{Menon2013,Menon2019,Lauer2019,Crawford2020,Munson2022} have shown good agreement between the nucleosynthesis and the observed abundances in RCB atmospheres. 
Furthermore, the inferred number of galactic RCB stars \citep[e.g.,][]{Tisserand2020} is broadly consistent with that expected from the He WD - CO WD merger rate \citep[e.g.,][]{Ruiter2009,Karakas2015}, whereas the final flash stars spend too little time in the RCB phase to explain the number of known RCB stars \citep[][]{Clayton2006}. 
It has been suggested that dLHdC stars are similarly formed through mergers, but with different mass ratios \citep[e.g.,][]{Karambelkar2022,Tisserand2022}.

Ground-based studies of RCB stars show semiregular variability with photometric semi-amplitudes between 0.05 mag and 0.3 mag and radial velocity semi-amplitudes between 10 and 20 km s$^{-1}$ \citep[e.g.,][]{Alexander1972,Lawson1990,Lawson1994,Lawson1997,Alcock2001,Percy2004,Percy2023}. 
While \cite{Alexander1972} and \cite{Lawson1997} suggest that the variability is due to radial pulsations, \cite{Feast2019} find no coherent periodicity in the long-term radial velocity observations of R CrB itself, and suggest that the variability is instead due to large surface convective plumes. 
However, \cite{Alcock2001} were able to extract periods for several HdC stars using MACHO data with a long base-line from 1992 to 1999 (e.g., star 18.3325.148 with 83.8 days, star 12.10803.56 with 50.5 days). 
Distinguishing between the effects of pulsations and surface convective plumes is an interesting observational question, but for the purpose of this paper, we proceed by interpreting periodic HdC variability as pulsations. The pulsation periods range from 20 days to $>100$ days, but among 14 RCB stars studied by \cite{Percy2023} using data from the All-Sky Automated Search for Supernovae \citep[ASAS-SN;][]{Shappee2014,Kochanek2017}, the periods cluster between 30 and 50 days. Similar periods are found in the infrared light curves of 5 RCB stars \citep{Karambelkar2021}. 
Only 13 out of 32 dLHdC stars have detectable variability from ASAS-SN \citep[][]{Tisserand2022}, with lower peak-to-peak photometric amplitudes of $0.05-0.15$ mag and shorter periods of $10-40$ days , and lower radial velocity amplitudes \citep[][]{Lawson1997,Tisserand2023}. 

Previous nonadiabatic calculations have suggested that the strange mode instability \citep[for a review, see][]{Saio2009}, which exists in highly nonadiabatic and radiation pressure-dominated environments, could excite these pulsations both for radial \citep[e.g.,][]{Saio1984,Gautschy1990,Saio1998,Saio2008,GYREIII} and nonradial modes \citep[][]{Glatzel1992,Saio2008}. 

We argue here that HdC pulsations are solar-like oscillations \citep[for a review of solar-like oscillations, see, e.g.,][]{Chaplin2013}, which are acoustic oscillations ($p$-modes) stochastically excited by surface convection \citep[see e.g., ][ for discussions on the excitation and damping mechanism]{Goldreich1977,Goldreich1994,Samadi2001,Belkacem2006}, unlike self-excited pulsations driven by instabilities like the kappa mechanism. 
First discovered in the Sun, solar-like oscillations exist in all subgiants and red giants up to and beyond the tip of the red giant branch \citep[e.g.,][]{Mosser2013,Stello2014,Yu2020}. These discoveries were possible due to accurate photometric measurements enabled by space-based missions such as CoRoT \citep{Baglin2006,deRidder2009} and Kepler \citep[][]{Borucki2010,Beck2011}. 
Discovery of solar-like oscillations has allowed for the asteroseismological derivation of a star's mass and radius, as well as probes into its core structure \citep[e.g.,][]{Bedding2011,Mosser2012_core}, core rotation rate \citep[e.g.,][]{Beck2012,Mosser2012_spin}, and internal magnetic fields \citep[e.g.,][]{Fuller2015,Stello2016}.

The power spectrum of solar-like oscillations consists of many sharp individual peaks whose heights follow a Gaussian-like envelope, with a frequency of maximum power, $\numax$. Modes of the same angular degree but neighboring radial orders are separated by the large frequency separation, $\Delta \nu$. The former scales as the acoustic cutoff frequency, 
$\numax \propto \nu_{\mathrm{ac}} \propto \cs/H$ \citep[e.g.,][]{Kjeldsen1995}, where $\cs = \sqrt{\Gamma_{1} P / \rho}$ is the adiabatic sound speed, $H = P/(\rho g) $ is the pressure scale height, and $\Gamma_{1}$ is the first adiabatic exponent. 

HdC stars have neutral He-dominated atmospheres (mean molecular weight $\mu\approx4$), as opposed to $\mu\approx1.3$ for H-dominated stars. Since $\cs\propto\sqrt{1/\mu}$ and $ H \propto P \propto 1/\mu$ for an ideal gas \citep[e.g.,][]{Yildiz2016,Viani2017}, we expect a correction factor $\sqrt{\mu/\mu_{\odot}} \approx \sqrt{4/1.3} \approx 1.75$ for the solar-scaled $\numax$ relation when applied to HdC stars:
\begin{equation}
\label{equ:numax}
    \numax
    = 
     0.41\,\mu\mathrm{Hz} \left( \frac{M}{0.7\,\msun} \right) \left( \frac{R}{100\,\rsun} \right)^{-2} \left( \frac{\Teff}{5000\,\mathrm{K}} \right)^{-1/2}. 
\end{equation}
which corresponds to a period of $30$ d, in agreement with the observed period range for HdC stars. 

For high radial order $p$-modes, the large frequency separation is given by 
\begin{equation}
\label{equ:deltanuintegral}
    \Delta \nu \approx \left( 2 \int \frac{dr}{\cs} \right)^{-1} , 
\end{equation}
which for our models yield, to within 20\%, 
\begin{equation}
\label{equ:deltanunew}
    \Delta \nu
    = 0.18\,\mu\mathrm{Hz} \left( \frac{M}{0.7\,\msun} \right)^{1/2} \left( \frac{R}{100 \rsun} \right)^{-3/2} ,
\end{equation}
which is about $1.6$ times the solar relation\footnote{For our fiducial model, eqn \ref{equ:deltanuintegral} is about $\approx 1.4$ times the solar relation at $\Teff\lesssim5000\,\mathrm{K}$, and increases to $\approx 1.8$ at higher $\Teff$ as $L$ increases. }. 

These simple estimates imply that HdC stars would exhibit solar-like oscillations in the period range that is observed and with radial orders of $n=2-3$. 
Even at these low radial orders, we find that $\nu_{n+1} - \nu_{n}$ agrees with the asymptotic relation (eqn \ref{equ:deltanuintegral}) within 10\% for $n \geqslant 2$. Therefore, the use of $\Delta \nu$ to assign radial orders is still applicable. 

We present evolutionary models and pulsation calculations for HdC stars to show that HdC could exhibit solar-like oscillations in the observed period range. 
In Sec \ref{sec:model}, we construct 3 HdC star models with a $0.50-0.55\,\msun$ CO core and a $0.20-0.25\,\msun$ He-rich envelope. 
We then present adiabatic calculations for our fiducial $0.20+0.50\,\msun$ model in Sec \ref{sec:fiducial} and other models in Sec \ref{sec:others}. 
Non-adiabatic calculations are summarized in Sec \ref{sec:nad} and detailed in Appendix \ref{sec:nad_appendix}. We conclude in Sec \ref{sec:conclusion} by comparing to observations and highlighting future work. 

\section{H Deficient Stellar models} 
\label{sec:model}

We prepare 3 non-rotating HdC models following \cite{Lauer2019,Schwab2019,Crawford2020,Munson2022}, using the Modules for Experiments in Stellar Astrophysics \citep[$\mesa$  version $\tt r22.11.1$; ][]{MESAI,MESAII,MESAIII,MESAIV,MESAV,MESAVI}. 
We adopt the \cite{Asplund2000} compositions for the majority RCB population, with $Y=0.937$, $X(^{12}\mathrm{C})=0.041$, $X(^{14}\mathrm{N})=0.015$,  and $X(^{16}\mathrm{O})=X(^{18}\mathrm{O})=0.0035$. This implicitly assumes that chemical mixing between the core and envelope has occurred during and immediately after the merger, and so we do not include chemical mixing processes, unlike the aforementioned works. We adopt the $\tt AESOPUS$ \citep{Marigo2009} low-temperature opacity tables for varying CNO abundances. 
We adopt a mixing length parameter of $\alphaMLT=2$ and a metallicity of $Z=0.006$. 

Our models mimic post-merger objects where the CO core has cooled after completing He burning. We accrete an envelope with the \cite{Asplund2000} composition onto a CO core with central temperature $2 \times 10^{7}\,\mathrm{K}$, and then inject entropy into the envelope to allow it to reach temperatures for He shell-burning and to expand to sizes typical for an HdC star. 
We experiment with different mass combinations, with $\mhe+\mco=0.20+0.50$ and $0.25+0.55\,\msun$, and the effect of including wind mass loss \citep[][with efficiency factor $\eta = 0.02$]{Bloecker1995}.

We perform adiabatic calculations with $\gyre$ \citep[version $\tt 6.0$; ][]{GYREI,GYREII,GYREIII}, for angular degrees $l=0,1,2$. 
The outer surface boundary condition is taken to be zero surface pressure. 
For most models, we excise the degenerate WD core, which eliminates mixed gravito-acoustic modes deep in the core and reduces computation cost. We take the inner boundary condition to be zero radial displacement. For a few models we explore the mixed gravito-acoustic modes with a full stellar model with a regularity inner boundary condition. 
In the following, we refer to the radial order $n_{\mathrm{pg}}$ as defined in \cite{Takata2006}. For $l>0$, $n_{\mathrm{pg}}$ refers to that of the excised-core model, and not the full stellar model. 
We explore the non-adiabatic modes using the contour method \citep{GYREIII} in Sec \ref{sec:nad} and Appendix \ref{sec:nad_appendix}. 

Our model names start with (initial He mass)+(initial CO mass), and if mass loss is involved, ``ML'' is added to the end. 
For example, the 0.20+0.50ML model has initially $\mhe=0.20\,\msun$ and $\mco=0.50\,\msun$ and has a \cite{Bloecker1995} wind mass loss.

Our $\mesa$ and $\gyre$ input and output files are available at Zenodo (\url{https://doi.org/10.5281/zenodo.10139856}). 

\section{Asteroseismic results}

\subsection{Adiabatic fiducial model}
\label{sec:fiducial}

\begin{figure*}[t!]
\centering
\fig{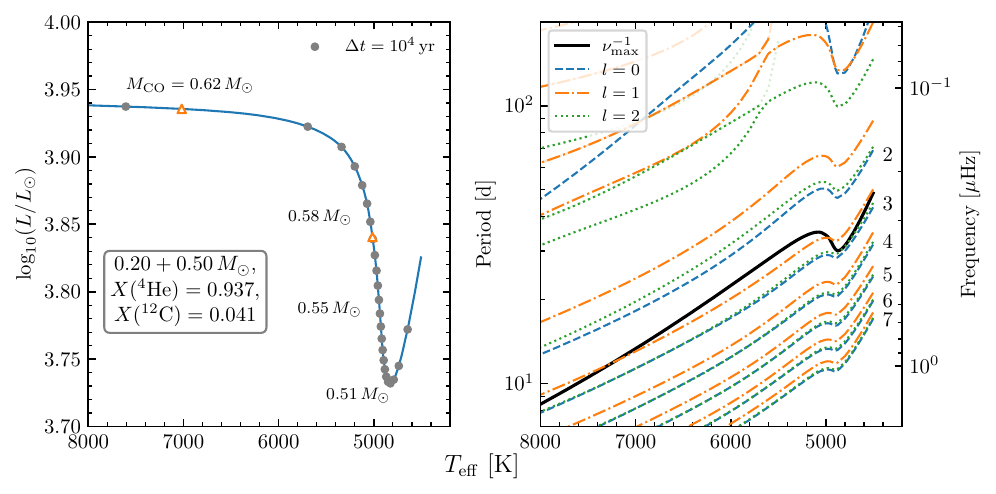}{ 0.8 \textwidth }{}
\caption{ 
Evolution on the HR diagram (left) and adiabatic mode periods and frequencies (right) of model $0.20+0.50$. \textbf{Left}: Circle markers on the HR diagram indicate time intervals of $10^{4}$ yrs, with a few labeled with the mass of the CO core. Two models (orange triangles) are chosen for detailed examination in Figure \ref{fig:fig2}. 
\textbf{Right}: Mode periods for $l=0$ (blue dashed), $l=1$ (orange dot-dashed), and $l=2$ (green dotted). The period corresponding to $\numax$ (Eqn \ref{equ:numax}) is shown as a black solid line. For $l=0$ modes we label their $n_{\mathrm{pg}}$ number, and for $l=1,2$, modes with large mode inertia are shown as semi-transparent lines. 
\label{fig:fig1}}
\end{figure*}

We first discuss model $0.20+0.50$ as our fiducial. 
The left panel of Figure \ref{fig:fig1} shows the evolution of this model on the Hertzsprung-Russell (HR) diagram. 
We only exhibit the long-lived phase of stable He shell burning. 
As He shell burning proceeds, the core mass grows to $\approx 0.6\,\msun$ at a rate of $\approx 10^{-6}\,\msunyr$ appropriate for steady He burning, and the luminosity increases. 
This and other models follow a core mass-luminosity relation given by $\log_{10} (L/\lsun)  = 2.71 +  \mco/(0.5\,\msun) $
for $0.5 \lesssim \mco \lesssim 0.8\,\msun$ to within 20\%. The variance of 20\% is due to varying initial CO core masses \citep[e.g., ][]{Saio2002}. 
As the envelope becomes exhausted, the model moves to higher $\Teff$ toward the WD track. The mode periods of model $0.20+0.50$ during the evolution from low to high $\Teff$ are shown in the right panel.

In general, $\numax$ is close in frequency to the $n_{\mathrm{pg}}=3$ modes, and we may observe a couple of modes (few $\Delta \nu$) around $\numax$. As expected for higher radial orders, eqn \ref{equ:deltanuintegral} holds within $\approx 10\%$ for $n_{\mathrm{pg}} \geqslant 2$. It underpredicts $\Delta \nu_{12}$ by $5-20\%$ for $l=0$. 
Eqn \ref{equ:deltanunew}, which is $1.6$ times the solar relation, works for HdC stars to within 20\% for all models. 

\begin{figure*}[t!]
\centering
\fig{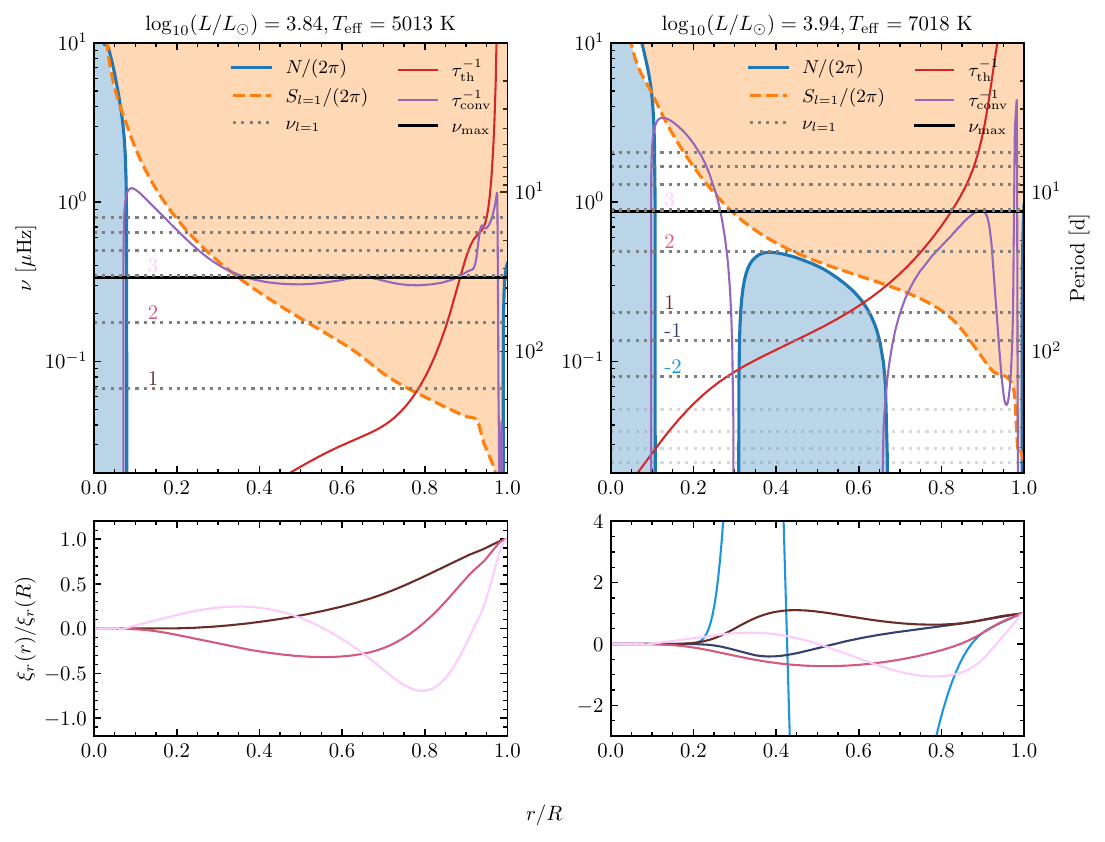}{ 0.8 \textwidth }{}
\caption{ 
Mode propagation diagram for $l=1$ modes (top) and radial displacement eigenfunctions for a few selected modes (bottom) for the two models selected in Figure \ref{fig:fig1}. 
$r/R=0$ refers to the center of the star. 
\textbf{Top}: Blue solid and orange dashed lines show the Brunt-V\"{a}is\"{a}l\"{a} and Lamb frequencies. Blue and orange shaded regions correspond to $g$- and $p$-mode propagation regions; whereas the white region corresponds to an evanescent zone. Red and purple solid lines correspond to the thermal and convective eddy turnover timescales (Eqns \ref{equ:tauconv} \& \ref{equ:tauth}). Black solid line shows $\numax$ (Eqn \ref{equ:numax}). Grey dotted lines indicate $l=1$ mode frequencies found by the excised-core model. Modes with large mode inertia are shown as semi-transparent lines in the right panel. A few selected excised-core modes are labelled by their $n_{\mathrm{pg}}$ numbers, and the corresponding radial displacement eigenfunctions of the full stellar model (with corresponding colors) are shown in the bottom. 
\label{fig:fig2}}
\end{figure*}

Mode propagation diagrams for two selected models are shown in Figure \ref{fig:fig2}. 
Acoustic $p$-modes have $\omega^{2} > N^{2}$ and $\omega^{2} > S^{2}_{l}$, where $N^{2}$ and $S^{2}_{l} = l (l+1) \cs^{2} / r^{2}$ are the Brunt-V\"{a}is\"{a}l\"{a} and Lamb frequencies squared, and propagate in the orange region. Gravity $g$-modes have $\omega^{2} < N^{2}$ and $\omega^{2} < S^{2}_{l}$, and propagate in the blue region. Elsewhere, the modes are evanescent. Note that the blue and white regions correspond to radiative ($N^{2}>0$) and convective zones ($N^{2}<0$). As the $p$-mode and $g$-mode propagation regions overlap in frequency space, we expect modes with mixed characters from both \citep[e.g.,][]{Osaki1975,Aizenman1977,Christensen2004,Dupret2009}. 

At low $\Teff \lesssim 5400\,\mathrm{K}$ (left panel), the envelope is fully convective, whereas at high $\Teff$ (right panel), the envelope shows two convective zones separated by a radiative zone. The deep convection zone is associated with the iron bump opacity peak, and the outer with the helium ionization zones. This radiative zone creates a $g$-mode cavity in addition to that in the non-convecting core, thus creating mixed modes even when the CO core is excised. Some of the mixed modes have predominantly $g$-mode character. If we examine their mode inertia, given by \citep[e.g.,][]{Aerts2010,Dupret2009} 
\begin{equation}
\label{equ:inertia}
    E_{\mathrm{mode}} = \frac{ \int^{R}_{0} \left[ \xi^{2}_{r}(r) + l (l+1) \xi^{2}_{h}(r) \right] 4 \pi r^{2} \rho(r) d r }{ M \left[ \xi^{2}_{r}(R) + l (l+1) \xi^{2}_{h}(R) \right] }, 
\end{equation}
where $\xi_{h}$ and $\xi_{r}$ are the horizontal and radial displacement eigenfunctions, we find that these $g$-dominated modes have much larger mode inertia than $p$-dominated modes, since these $g$-dominated modes have relatively larger amplitudes deep in the envelope where $\rho$ is higher. 
We show these $g$-dominated modes as semitransparent lines in Figures \ref{fig:fig1} \& \ref{fig:fig2}. We note that the envelope radiative zone separating the two convection zones grows in extent with increasing $\Teff$. Detection of periods of $\gtrsim 100$ days \citep[as found in e.g.,][]{Lawson1990,Rao2015,Karambelkar2021} would provide useful information on the envelope structure of HdC stars; such $\gtrsim 100$-d periods have been suggested by \cite{Lawson1990} to be non-radial pulsations as these periods appear longer than that predicted for the radial fundamental mode.

\begin{figure*}[t!]
\centering
\gridline{
\fig{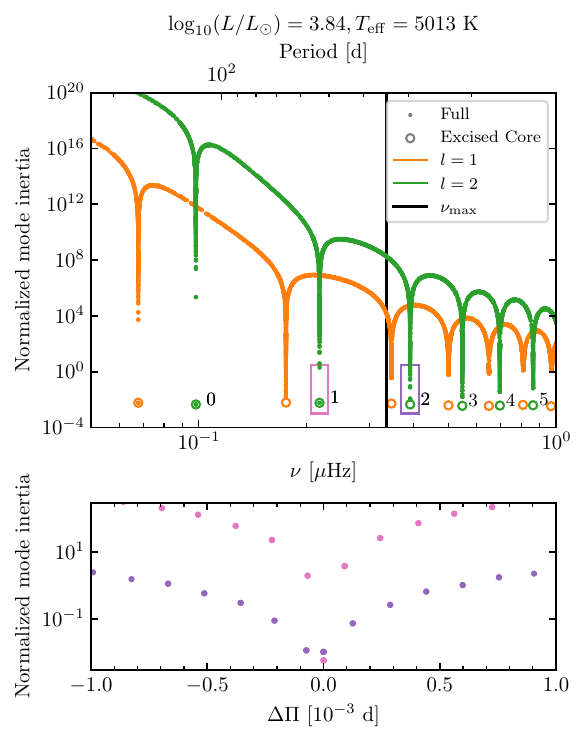}{0.5\textwidth}{}
\fig{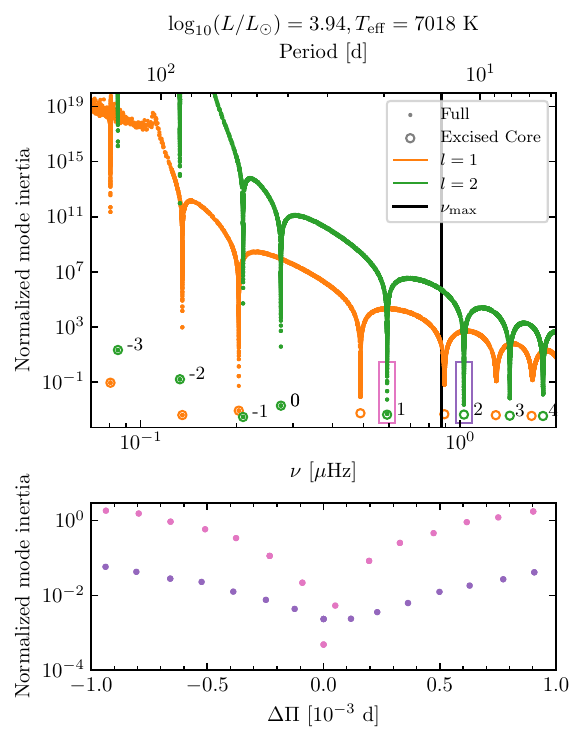}{0.5\textwidth}{}
}
\caption{ 
Normalized mode inertia for the two models selected in Figure 1. 
Dots show modes recovered in the full stellar model (note that these are discrete points), and open circles show those recovered in the excised-core model, with orange (green) colors corresponding to $l=1(2)$. 
Bottom panel shows the period spacing relative to the $l=2,n_{\mathrm{pg}}=1,2$ mode (pink and purple respectively), illustrating strong core-envelope coupling. 
\label{fig:fig3}}
\end{figure*}

The non-convecting WD core creates another $g$-mode cavity, allowing core $g$-modes to couple to the envelope $p$-modes. We calculate $l=1 , 2$ modes for the full stellar model corresponding to the models in Figure \ref{fig:fig2}. These modes are shown as orange and green dots in the top panel of Figure \ref{fig:fig3}. Each local minimum in mode inertia corresponds to a $p$-dominated mixed mode, and agrees well with the excised-core modes which are shown as orange circles. Furthermore, the $p$-dominated modes have identical eigenfunctions in the envelope between the excised-core models and full models (the latter are shown in the bottom panel of Figure \ref{fig:fig2}). The successful recovery of the $p$-dominated mixed modes supports our approach to perform most calculations with excised-core models. 

For low frequency $p$-dominated modes, the mode inertia agrees well between the full model and excised core model. 
However, as the frequency increases, the mode inertia of the full models increase relative to the excised-core models. The reason is that the evanescent region shrinks (white region in top panels of Figure \ref{fig:fig2}) as the frequency increases, allowing for higher amplitude of the mode eigenfunctions near the core. The higher density there increases the mode inertia (see Eqn \ref{equ:inertia}). 
This effect is stronger for $l=1$ modes than $l=2$, because the Lamb frequency squared $S^{2}_{l} \propto l (l+1) $ is higher for $l=2$, which widens the evanescent region and weakens mode coupling for $l=2$. 

The bottom panels of Fig \ref{fig:fig3} show the difference in mode periods relative to the $p$-dominated mixed modes, for $l=1$, $n_{\mathrm{pg}}=3,4$. 
Away from the $p$-dominated mixed mode, the period spacing $\Delta \Pi$ is regular and agrees well with $2 \pi^{2}/ [\sqrt{(l(l+1))} \int N^{2} d\ln r ]$, reflecting the asymptotic behavior of the core $g$-modes which have $n \sim 10,000$. In this model, we removed by hand a spike in $N^{2}$ coming from an artificial composition discontinuity between the old CO core and the fresh CO from He shell burning , because we assumed C/O=50/50 for the core during initial model construction, but realistic burning produces fresh C/O$\approx60/40$; mixing during and after the merger, which we do not model, could smooth this composition discontinuity. Otherwise, the period spacing would not agree with the asymptotic relation, due to this buoyancy glitch \citep[e.g.,][]{Cunha2015}. 
Resolving any of the $g$-dominated mixed modes could offer deep insights into the core structure of HdC stars as has been done for Kepler red giants \citep[e.g.,][]{Bedding2011,Mosser2012_core}, but is a challenge given that $\Delta \Pi \approx 3 \times 10^{-4} $ days requires an observation time of $\approx 10$ yrs.

Our calculations do not capture the coupling between convection and oscillations. 
The convective eddy turnover timescale, 
\begin{equation}
\label{equ:tauconv}
    \tau_{\mathrm{conv}} = \frac{ \alphaMLT H }{ v_{\mathrm{c}} } ,
\end{equation}
where $v_{\mathrm{c}}$ is the convective speed, is comparable to the periods of the $n_{\mathrm{pg}}=2,3$ modes. This suggests that convection could play an important role for these modes. 
Furthermore, the local thermal timescale, 
\begin{equation}
\label{equ:tauth}
    \tauth(m) = \frac{ \int^{M}_{m} \cp T d m' }{ L } ,
\end{equation}
where $\cp$ is the heat capacity at constant pressure, is shorter than the mode periods over a large fraction of the stellar envelope, suggesting that non-adiabatic effects may be important, as we discuss in Sec \ref{sec:nad} and Appendix \ref{sec:nad_appendix}.

\subsection{Other models}
\label{sec:others}

\begin{figure*}[]
\centering
\fig{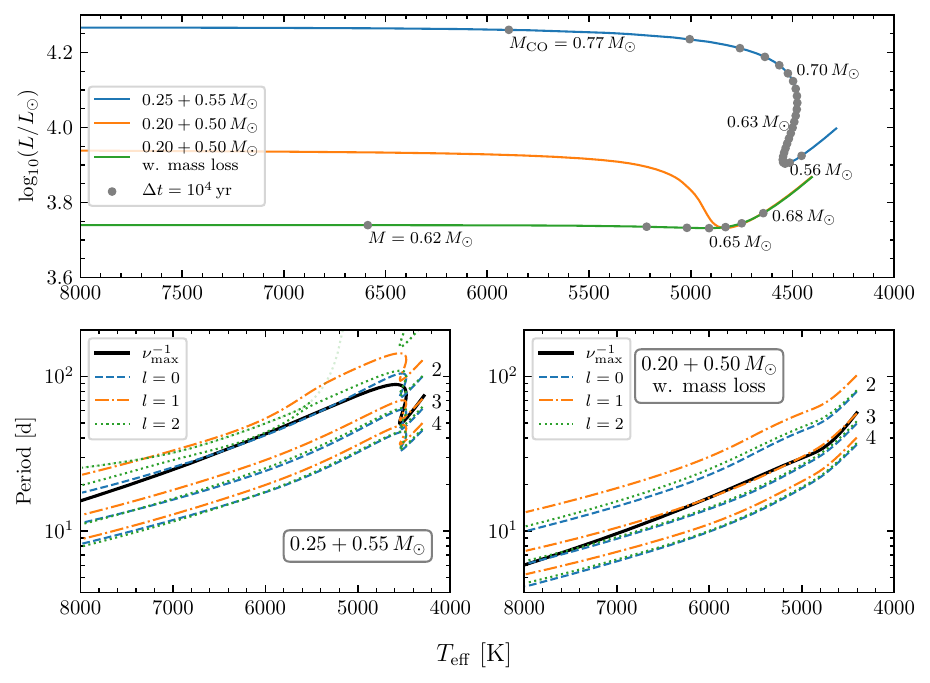}{ 0.8 \textwidth }{}
\caption{ 
Evolution on the HR diagram (top) and adiabatic mode periods (bottom) of the $0.25+0.55\,\msun$ (left) and $0.20+0.50\,\msun$ merger models with mass loss (right). 
Symbols carry the same meaning as Figure \ref{fig:fig1}. 
\textbf{Top}: Models shown are $0.25+0.55\,\msun$ (blue), $0.20+0.50\msun$ (orange) and $0.20+0.50\,\msun$ with mass loss (green). We show time intervals of $10^{4}$ yrs (grey points) only for the first and last models, and for the $0.20+0.50\,\msun$ model with mass loss, we show the total mass instead of CO core mass. 
\label{fig:fig4}}
\end{figure*}

We explore the parameter space occupied by HdC stars by introducing models $0.25+0.55$ and $0.20+0.50$ML. 
The HR diagrams of these two models are shown in the top row of Figure \ref{fig:fig4}. 
Model $0.25+0.55$ (blue) represents a higher mass and higher luminosity HdC star ($\log_{10}(L/\lsun)\approx 4.26$) and has a lifetime of $\approx 2.5\times10^{5}$ yr. 
For model $0.20+0.50$ML (green), rapid mass loss decreases its mass down to $0.65\,\msun$ once it reaches the HdC stage, accompanied by a drop in luminosity to $\log_{10}(L/\lsun)\approx 3.74$. It then evolves to the blue at roughly constant luminosity as its mass decreases to $\approx 0.62\,\msun$. 
Its lifetime in the HdC stage ($\approx 7 \times 10^{4}$ yr) is shorter than without mass loss ($\approx 2.5 \times 10^{5}$ yr). 
This model represents a lower mass and lower luminosity HdC star, compared to model $0.20+0.50$ which has no mass loss ($\log_{10}(L/\lsun)\approx 3.95$). 
Note that, as explored by \cite{Schwab2019}, the RCrB lifetime, total mass, and luminosity reached by the models sensitively depend on the mass loss prescription, with higher mass loss generally yielding a shorter lifetime, lower mass and lower luminosity.

Mode periods near $\numax$ are shown in the bottom row of Figure \ref{fig:fig4}. 
Model $0.25+0.55$ (left panel) shows longer periods than model $0.20+0.50$, and model $0.20+0.50$ML (right) shows shorter periods, since the global dynamical timescale $\tdyn \propto \sqrt{ R^{3} / M } $ increases with radius. 
Because of its scaling as $\propto R^{2}$, the period corresponding to $\numax$ shows a similar trend, with $\numax$ closest to the $n_{\mathrm{pg}}=2$ mode in model $0.25+0.55$, and $n_{\mathrm{pg}}=3$ mode in model $0.20+0.50$ML.

\subsection{Summary of non-adiabatic effects}
\label{sec:nad}

We investigate non-adiabatic effects for $l=0$ modes in Appendix \ref{sec:nad_appendix}. 
Unlike previous works \citep[e.g.,][]{Wood1967,Weiss1987_pulsation,Gautschy1990,Saio1998}, our focus is not on the mechanism for self-excitation, but we are interested in how non-adiabaticity changes the real part of the frequency, particularly near $\numax$. 
For models $0.20+0.50$ and $0.20+0.50$ML, the $n_{\mathrm{pg}}=2-4$ mode frequencies near $\numax$ do not deviate from their adiabatic counterparts by more than $20\%$. At $\Teff\lesssim5500\,\mathrm{K}$ where non-adiabatic effects are weak, the agreement is within 2\%. 
Model $0.25+0.55$ behaves similarly at $\Teff\lesssim5500\,\mathrm{K}$. However, at higher $\Teff$ additional modes appear, and correspondence between the non-adiabatic and adiabatic modes becomes complicated.

\section{Conclusion}
\label{sec:conclusion}

In this work, we argue that HdC pulsations are, by nature, solar-like oscillations resulting from a vigorously convecting envelope. We use the solar-scaled $\numax$ relation to He-rich HdC stars, with a $\mu$ correction for He-rich compositions (Eqn \ref{equ:numax}) to focus our work on those modes most likely of large amplitude.
We construct $\mesa$ models with a $0.5-0.55\,\msun$ CO core and $0.2-0.25\,\msun$ He-rich envelopes, and perform adiabatic calculations for $l=0,1,2$ using $\gyre$. 
We find that $\Delta\nu$ is generally $1.4-2$ times the solar relation, and suggest a new relation (Eqn \ref{equ:deltanunew}) that is $\approx1.6$ times the solar relation and applies within 20\% for He-rich objects. 
Modes near $\numax$ have radial orders $n_{\mathrm{pg}}=2-3$, which means a few modes are excited for HdC stars, in agreement with observations that only one or two periods are observed for HdC stars \citep[e.g.,][]{Karambelkar2021,Percy2023}. 

Our $\numax$ corrected for He-rich compositions matches the $30-50$ d periods typically found in RCB stars \citep[][]{Percy2023}. 
As HdC star models move to higher $\Teff\gtrsim5400\,\mathrm{K}$, a radiative zone appears in the otherwise fully convective envelope, which creates a $g$-mode cavity and may explain some of the longer periods ($\approx 100$ days) observed in HdC stars \citep[e.g.,][]{Lawson1990}. 
As shown in Sec \ref{sec:others}, a lower mass HdC star has a lower luminosity due to its smaller core mass. Because of its smaller radius, its pulsation periods (and period corresponding to $\numax$) are also shorter. 
This points to a smaller mass for dLHdC stars, which have shorter pulsation periods as well as lower absolute magnitudes than RCB stars \citep[$\approx 10-40$ days; ][]{Tisserand2022}. 

Space-based observations of solar-like oscillations in sub-giants and giants 
find a strong relation between the semi-amplitude of pulsations with the stellar luminosity, 
extending all the way up the red giant branch \citep[e.g.,][]{Huber2011,Yu2018,Yu2020}. 
At the brightest end (e.g. at the tip of the RGB and beyond into the AGB)
these are fully convective, H-rich stars in the realms of the ground based studied 
long-period variables \citep[e.g.,][]{Mosser2013,Stello2014,Yu2020}. 
Though we have little available phenomenology for the HdC stars, their 
observed amplitudes ($0.05-0.3$ mag) are consistent with comparably luminous H-rich counterparts \citep[$\approx0.1$ mag; ][]{Yu2020}, 
giving us additional confidence in our solar-like oscillation hypothesis.
Whether the tendency of the observed modes in these luminous stars to be 
preferentially dipolar \citep[][]{Stello2014} also proves true in  the HdC realm remains to be seen. 
HdC stars also have RV variability with peak-to-peak amplitude $\approx 10 - 20\,\mathrm{km}\,\mathrm{s}^{-1}$ \citep[e.g.,][]{Lawson1997,Tisserand2023}. If interpreted as radial pulsations \citep[however, see][]{Feast2019}, this would translate to a radial amplitude of $\delta R \approx \delta v / \omega \approx  (10\,\mathrm{km}\,\mathrm{s}^{-1})/(2 \pi / 40\,\mathrm{d}) \approx 8\,\rsun$, which seems reasonable given that this is around 10\% of the stellar radius. 

In addition, dLHdC stars have lower absolute magnitudes than RCB stars with the same color and shorter pulsation periods\citep[][]{Tisserand2022}, which as our models show, hints that dLHdC stars may have lower masses. 
If true, then dLHdC stars are ideal for studying pulsations in HdC stars, not just because they do not show dust ejection episodes, but also because non-adiabatic effects are less important. 

Future photometric studies of HdC stars using e.g., TESS \citep[][]{Ricker2014},  Palomar Gattini IR \citep[][]{De2020} or Plato \citep[][]{Rauer2014}, may uncover more pulsation periods for HdC stars. 
For example, due to revisits, TESS can observe some HdC stars over multiple pulsation periods. dLHdC stars are ideal targets for TESS since they show no declines and have shorter pulsation periods \citep[$\approx10-40$ days;][]{Tisserand2022} which is shorter than or comparable to TESS observation of a single sector (27 days). 
Determination of $\numax$ and $\Delta \nu$ could then yield asteroseismic masses and radii for HdC stars. 

We also perform non-adiabatic calculations for $l=0$. We find that at $\Teff \lesssim 5500\,\mathrm{K}$, the mode frequencies do not deviate much from their adiabatic counterparts, particularly near $\numax$. This remains true at higher $\Teff$ for models $0.20+0.50$ and $0.20+0.50$ML, but additional modes appear for model $0.25+0.55$ which complicates mode identification. Nevertheless, most HdC stars studied by \cite{Crawford2023} have $\Teff \lesssim 6500\,\mathrm{K}$, which means non-adiabatic effects are likely unimportant for them. 

Finally, these 1D models have envelopes that are significantly radiation pressure-dominated, with radiation pressure accounting for $40-80$\% of the total pressure. At the opacity peaks associated with the iron bump and the two helium ionization zones, the local luminosity is super-Eddington, creating a density inversion at the first helium ionization zone. Convection is likely inefficient in the sense of strong radiative losses. A 3D radiation hydrodynamic simulation similar to \cite{Jiang2018} will provide much deeper insights into the envelope structure, and may furthermore clarify whether HdC variability is due to pulsations or surface convective plumes \citep[e.g.,][]{Feast2019}.

\begin{acknowledgements}

We thank the referees for their constructive suggestions that have greatly improved our manuscript. 
We thank Richard Townsend for helpful conversations about running $\gyre$, and Viraj Karambelkar and May Pedersen for discussions of the observational impacts of our work. 
We thank Kris Stanek for pointing us to \cite{Percy2023} on HdC observations using ASAS-SN. We are grateful to Patrick Tisserand  for discussions on HdC observations. 
This work was supported, in part, by the National Science Foundation through grant PHY-1748958, and by 
the Gordon and Betty Moore Foundation through grant GBMF5076. 
Use was made of computational facilities purchased with funds from the National Science Foundation (CNS-1725797) and administered by the Center for Scientific Computing (CSC). The CSC is supported by the California NanoSystems Institute and the Materials Research Science and Engineering Center (MRSEC; NSF DMR 2308708) at UC Santa Barbara.

\end{acknowledgements}

\software{
\texttt{MESA} \citep[v22.11.1;][]{MESAI,MESAII,MESAIII,MESAIV,MESAV,MESAVI}, \texttt{GYRE} \citep[v6.0;][]{GYREI,GYREII,GYREIII}, 
\texttt{py\_mesa\_reader} \citep{bill_wolf_2017_826958},
\texttt{ipython/jupyter} \citep{perez_2007_aa,kluyver_2016_aa},
\texttt{matplotlib} \citep{hunter_2007_aa},
\texttt{NumPy} \citep{numpy2020}, 
\texttt{SciPy} \citep{scipy2020}, 
\texttt{Astropy} \citep{astropy:2013,astropy:2018},
and 
\texttt{Python} from \href{https://www.python.org}{python.org}
}


\clearpage

\appendix

\twocolumngrid

\section{Non-adiabatic effects}
\label{sec:nad_appendix}

At low $\Teff$ ($\lesssim 5500\,\mathrm{K}$), the non-adiabatic modes show a clear correspondence with the adiabatic modes. The correspondence can be tracked on the complex frequency plane (i.e., $\mathrm{Re}(\nu)$ vs $\mathrm{Im}(\nu)$) by gradually scaling the local thermal timescale by a factor $\alpha_{\mathrm{th}}$, from a large number corresponding to the adiabatic limit, to $1$ corresponding to the non-adiabatic limit \citep[e.g.,][]{Saio1984,Gautschy1990}. 
The low radial order (low frequency) non-adiabatic modes, especially those near $\numax$, have real frequencies that agree well with their adiabatic counterparts, except for $n_{\mathrm{pg}}=1$ (see later discussion). 
For $n_{\mathrm{pg}}=2-4$, the agreement is within 2\%. 
Therefore, mode identification using adiabatic calculations should be sufficient for $l=0$ modes with $\mathrm{Re}(\nu) \lesssim \numax$. 
Higher radial order modes have increasingly lower real frequencies relative to their adiabatic counterparts. 
This is because as non-adiabaticity increases, the sound propagation time changes from the adiabatic limit to the isothermal limit, and hence becomes longer \citep[e.g.,][]{Zinn2023}. With increasing radial order, heat exchange between neighboring temperature perturbations increases, enhancing the influence of non-adiabaticity. 
For these high radial order modes, we find that non-adiabaticity increases the number of radial nodes, in agreement with \cite{GYREIII}.

The picture becomes complicated as $\Teff$ increases ($\gtrsim 5500\,\mathrm{K}$). 
We first discuss the $0.20+0.50\,\msun$ merger models with and without mass loss. 
Most non-adiabatic modes that have a correspondence with the adiabatic modes have $\mathrm{Re}(\nu)$ that significantly deviate from their adiabatic counterparts and may approach that of another mode. 
At very low frequencies (at or below that of the $n_{\mathrm{pg}}=1$ mode), a number of modes with high mode inertia and/or high damping ($\mathrm{Im}(\nu)/\mathrm{Re}(\nu) \ll 0$) appear, which we speculate are secular in origin. In addition, we find a few damped modes that cross numerous other modes on a modal diagram (e.g., $\mathrm{Re}(\nu)$ vs $\Teff$). These modes rapidly decrease in $\mathrm{Re}(\nu)$ with $\Teff$ unlike the other modes, and we interpret these as the strange modes studied by \cite{Wood1967,Cox1980,Saio1984}. We do not further investigate the nature of these additional modes. Our present focus is on modes near $\numax$. Intriguingly, the $n_{\mathrm{pg}}=3$ non-adiabatic mode which is nearest to $\numax$, shows $\lesssim 8\%$ deviation from its adiabatic counterpart. The $n_{\mathrm{pg}}=1,2$ non-adiabatic modes both increase in $\mathrm{Re}(\nu)$ relative to their adiabatic counterparts, and the $n_{\mathrm{pg}}=4$ mode shows a decrease. The relative change in $\mathrm{Re}(\nu)$ for $n_{\mathrm{pg}}=2,4$ is modest, increasing from $\lesssim 2\%$ at $\Teff \lesssim 5500\,\mathrm{K}$ to $\approx 20\%$ at $\Teff \approx 8000\,\mathrm{K}$. Meanwhile, $\mathrm{Re}(\nu)$ of the $n_{\mathrm{pg}}=1$ mode increases up to $2-3$ times starting from $\Teff \approx 5000\,\mathrm{K}$. For the $0.25+0.55\,\msun$ model, many additional modes appear, with comparable $\mathrm{Im}(\nu)$ and mode inertia as the modes which have an adiabatic counterpart, making mode identification complicated. 

In short, non-adiabatic effects are not significant for low $\Teff$ ($\lesssim 5500\,\mathrm{K}$), but at high $\Teff$ may change mode frequencies relative to their adiabatic counterparts or even introduce additional modes. However, mode frequencies near $\numax$ may still be well approximated by the adiabatic limit. 


\bibliography{mesa,rcrb,software}{}

\begin{thebibliography}{}
\expandafter\ifx\csname natexlab\endcsname\relax\def\natexlab#1{#1}\fi
\providecommand{\url}[1]{\href{#1}{#1}}
\providecommand{\dodoi}[1]{doi:~\href{http://doi.org/#1}{\nolinkurl{#1}}}
\providecommand{\doeprint}[1]{\href{http://ascl.net/#1}{\nolinkurl{http://ascl.net/#1}}}
\providecommand{\doarXiv}[1]{\href{https://arxiv.org/abs/#1}{\nolinkurl{https://arxiv.org/abs/#1}}}

\bibitem[{{Aerts} {et~al.}(2010){Aerts}, {Christensen-Dalsgaard}, \& {Kurtz}}]{Aerts2010}
{Aerts}, C., {Christensen-Dalsgaard}, J., \& {Kurtz}, D.~W. 2010, {Asteroseismology}, \dodoi{10.1007/978-1-4020-5803-5}

\bibitem[{{Aizenman} {et~al.}(1977){Aizenman}, {Smeyers}, \& {Weigert}}]{Aizenman1977}
{Aizenman}, M., {Smeyers}, P., \& {Weigert}, A. 1977, \aap, 58, 41

\bibitem[{{Alcock} {et~al.}(2001){Alcock}, {Allsman}, {Alves}, {Axelrod}, {Becker}, {Bennett}, {Clayton}, {Cook}, {Dalal}, {Drake}, {Freeman}, {Geha}, {Gordon}, {Griest}, {Kilkenny}, {Lehner}, {Marshall}, {Minniti}, {Misselt}, {Nelson}, {Peterson}, {Popowski}, {Pratt}, {Quinn}, {Stubbs}, {Sutherland}, {Tomaney}, {Vandehei}, \& {Welch}}]{Alcock2001}
{Alcock}, C., {Allsman}, R.~A., {Alves}, D.~R., {et~al.} 2001, \apj, 554, 298, \dodoi{10.1086/321369}

\bibitem[{{Alexander} {et~al.}(1972){Alexander}, {Andrews}, {Catchpole}, {Feast}, {Llyod Evans}, {Menzies}, {Wisse}, \& {Wisse}}]{Alexander1972}
{Alexander}, J.~B., {Andrews}, P.~J., {Catchpole}, R.~M., {et~al.} 1972, \mnras, 158, 305, \dodoi{10.1093/mnras/158.3.305}

\bibitem[{{Asplund} {et~al.}(2000){Asplund}, {Gustafsson}, {Lambert}, \& {Rao}}]{Asplund2000}
{Asplund}, M., {Gustafsson}, B., {Lambert}, D.~L., \& {Rao}, N.~K. 2000, \aap, 353, 287

\bibitem[{{Astropy Collaboration} {et~al.}(2013){Astropy Collaboration}, {Robitaille}, {Tollerud}, {Greenfield}, {Droettboom}, {Bray}, {Aldcroft}, {Davis}, {Ginsburg}, {Price-Whelan}, {Kerzendorf}, {Conley}, {Crighton}, {Barbary}, {Muna}, {Ferguson}, {Grollier}, {Parikh}, {Nair}, {Unther}, {Deil}, {Woillez}, {Conseil}, {Kramer}, {Turner}, {Singer}, {Fox}, {Weaver}, {Zabalza}, {Edwards}, {Azalee Bostroem}, {Burke}, {Casey}, {Crawford}, {Dencheva}, {Ely}, {Jenness}, {Labrie}, {Lim}, {Pierfederici}, {Pontzen}, {Ptak}, {Refsdal}, {Servillat}, \& {Streicher}}]{astropy:2013}
{Astropy Collaboration}, {Robitaille}, T.~P., {Tollerud}, E.~J., {et~al.} 2013, \aap, 558, A33, \dodoi{10.1051/0004-6361/201322068}

\bibitem[{{Astropy Collaboration} {et~al.}(2018){Astropy Collaboration}, {Price-Whelan}, {Sip{\H{o}}cz}, {G{\"u}nther}, {Lim}, {Crawford}, {Conseil}, {Shupe}, {Craig}, {Dencheva}, {Ginsburg}, {Vand erPlas}, {Bradley}, {P{\'e}rez-Su{\'a}rez}, {de Val-Borro}, {Aldcroft}, {Cruz}, {Robitaille}, {Tollerud}, {Ardelean}, {Babej}, {Bach}, {Bachetti}, {Bakanov}, {Bamford}, {Barentsen}, {Barmby}, {Baumbach}, {Berry}, {Biscani}, {Boquien}, {Bostroem}, {Bouma}, {Brammer}, {Bray}, {Breytenbach}, {Buddelmeijer}, {Burke}, {Calderone}, {Cano Rodr{\'\i}guez}, {Cara}, {Cardoso}, {Cheedella}, {Copin}, {Corrales}, {Crichton}, {D'Avella}, {Deil}, {Depagne}, {Dietrich}, {Donath}, {Droettboom}, {Earl}, {Erben}, {Fabbro}, {Ferreira}, {Finethy}, {Fox}, {Garrison}, {Gibbons}, {Goldstein}, {Gommers}, {Greco}, {Greenfield}, {Groener}, {Grollier}, {Hagen}, {Hirst}, {Homeier}, {Horton}, {Hosseinzadeh}, {Hu}, {Hunkeler}, {Ivezi{\'c}}, {Jain}, {Jenness}, {Kanarek}, {Kendrew}, {Kern}, {Kerzendorf}, {Khvalko}, {King}, {Kirkby}, {Kulkarni},
  {Kumar}, {Lee}, {Lenz}, {Littlefair}, {Ma}, {Macleod}, {Mastropietro}, {McCully}, {Montagnac}, {Morris}, {Mueller}, {Mumford}, {Muna}, {Murphy}, {Nelson}, {Nguyen}, {Ninan}, {N{\"o}the}, {Ogaz}, {Oh}, {Parejko}, {Parley}, {Pascual}, {Patil}, {Patil}, {Plunkett}, {Prochaska}, {Rastogi}, {Reddy Janga}, {Sabater}, {Sakurikar}, {Seifert}, {Sherbert}, {Sherwood-Taylor}, {Shih}, {Sick}, {Silbiger}, {Singanamalla}, {Singer}, {Sladen}, {Sooley}, {Sornarajah}, {Streicher}, {Teuben}, {Thomas}, {Tremblay}, {Turner}, {Terr{\'o}n}, {van Kerkwijk}, {de la Vega}, {Watkins}, {Weaver}, {Whitmore}, {Woillez}, {Zabalza}, \& {Astropy Contributors}}]{astropy:2018}
{Astropy Collaboration}, {Price-Whelan}, A.~M., {Sip{\H{o}}cz}, B.~M., {et~al.} 2018, \aj, 156, 123, \dodoi{10.3847/1538-3881/aabc4f}

\bibitem[{{Baglin} {et~al.}(2006){Baglin}, {Auvergne}, {Barge}, {Deleuil}, {Catala}, {Michel}, {Weiss}, \& {COROT Team}}]{Baglin2006}
{Baglin}, A., {Auvergne}, M., {Barge}, P., {et~al.} 2006, in ESA Special Publication, Vol. 1306, The CoRoT Mission Pre-Launch Status - Stellar Seismology and Planet Finding, ed. M.~{Fridlund}, A.~{Baglin}, J.~{Lochard}, \& L.~{Conroy}, 33

\bibitem[{{Beck} {et~al.}(2011){Beck}, {Bedding}, {Mosser}, {Stello}, {Garcia}, {Kallinger}, {Hekker}, {Elsworth}, {Frandsen}, {Carrier}, {De Ridder}, {Aerts}, {White}, {Huber}, {Dupret}, {Montalb{\'a}n}, {Miglio}, {Noels}, {Chaplin}, {Kjeldsen}, {Christensen-Dalsgaard}, {Gilliland}, {Brown}, {Kawaler}, {Mathur}, \& {Jenkins}}]{Beck2011}
{Beck}, P.~G., {Bedding}, T.~R., {Mosser}, B., {et~al.} 2011, Science, 332, 205, \dodoi{10.1126/science.1201939}

\bibitem[{{Beck} {et~al.}(2012){Beck}, {Montalban}, {Kallinger}, {De Ridder}, {Aerts}, {Garc{\'\i}a}, {Hekker}, {Dupret}, {Mosser}, {Eggenberger}, {Stello}, {Elsworth}, {Frandsen}, {Carrier}, {Hillen}, {Gruberbauer}, {Christensen-Dalsgaard}, {Miglio}, {Valentini}, {Bedding}, {Kjeldsen}, {Girouard}, {Hall}, \& {Ibrahim}}]{Beck2012}
{Beck}, P.~G., {Montalban}, J., {Kallinger}, T., {et~al.} 2012, \nat, 481, 55, \dodoi{10.1038/nature10612}

\bibitem[{{Bedding} {et~al.}(2011){Bedding}, {Mosser}, {Huber}, {Montalb{\'a}n}, {Beck}, {Christensen-Dalsgaard}, {Elsworth}, {Garc{\'\i}a}, {Miglio}, {Stello}, {White}, {De Ridder}, {Hekker}, {Aerts}, {Barban}, {Belkacem}, {Broomhall}, {Brown}, {Buzasi}, {Carrier}, {Chaplin}, {di Mauro}, {Dupret}, {Frandsen}, {Gilliland}, {Goupil}, {Jenkins}, {Kallinger}, {Kawaler}, {Kjeldsen}, {Mathur}, {Noels}, {Silva Aguirre}, \& {Ventura}}]{Bedding2011}
{Bedding}, T.~R., {Mosser}, B., {Huber}, D., {et~al.} 2011, \nat, 471, 608, \dodoi{10.1038/nature09935}

\bibitem[{{Belkacem} {et~al.}(2006){Belkacem}, {Samadi}, {Goupil}, {Kupka}, \& {Baudin}}]{Belkacem2006}
{Belkacem}, K., {Samadi}, R., {Goupil}, M.~J., {Kupka}, F., \& {Baudin}, F. 2006, \aap, 460, 183, \dodoi{10.1051/0004-6361:20065370}

\bibitem[{{Bloecker}(1995)}]{Bloecker1995}
{Bloecker}, T. 1995, \aap, 297, 727

\bibitem[{{Borucki} {et~al.}(2010){Borucki}, {Koch}, {Basri}, {Batalha}, {Brown}, {Caldwell}, {Caldwell}, {Christensen-Dalsgaard}, {Cochran}, {DeVore}, {Dunham}, {Dupree}, {Gautier}, {Geary}, {Gilliland}, {Gould}, {Howell}, {Jenkins}, {Kondo}, {Latham}, {Marcy}, {Meibom}, {Kjeldsen}, {Lissauer}, {Monet}, {Morrison}, {Sasselov}, {Tarter}, {Boss}, {Brownlee}, {Owen}, {Buzasi}, {Charbonneau}, {Doyle}, {Fortney}, {Ford}, {Holman}, {Seager}, {Steffen}, {Welsh}, {Rowe}, {Anderson}, {Buchhave}, {Ciardi}, {Walkowicz}, {Sherry}, {Horch}, {Isaacson}, {Everett}, {Fischer}, {Torres}, {Johnson}, {Endl}, {MacQueen}, {Bryson}, {Dotson}, {Haas}, {Kolodziejczak}, {Van Cleve}, {Chandrasekaran}, {Twicken}, {Quintana}, {Clarke}, {Allen}, {Li}, {Wu}, {Tenenbaum}, {Verner}, {Bruhweiler}, {Barnes}, \& {Prsa}}]{Borucki2010}
{Borucki}, W.~J., {Koch}, D., {Basri}, G., {et~al.} 2010, Science, 327, 977, \dodoi{10.1126/science.1185402}

\bibitem[{{Chaplin} \& {Miglio}(2013)}]{Chaplin2013}
{Chaplin}, W.~J., \& {Miglio}, A. 2013, \araa, 51, 353, \dodoi{10.1146/annurev-astro-082812-140938}

\bibitem[{{Christensen-Dalsgaard}(2004)}]{Christensen2004}
{Christensen-Dalsgaard}, J. 2004, \solphys, 220, 137, \dodoi{10.1023/B:SOLA.0000031392.43227.7d}

\bibitem[{{Clayton}(1996)}]{Clayton1996}
{Clayton}, G.~C. 1996, \pasp, 108, 225, \dodoi{10.1086/133715}

\bibitem[{{Clayton}(2012)}]{Clayton2012}
---. 2012, \jaavso, 40, 539, \dodoi{10.48550/arXiv.1206.3448}

\bibitem[{{Clayton} {et~al.}(2007){Clayton}, {Geballe}, {Herwig}, {Fryer}, \& {Asplund}}]{Clayton2007}
{Clayton}, G.~C., {Geballe}, T.~R., {Herwig}, F., {Fryer}, C., \& {Asplund}, M. 2007, \apj, 662, 1220, \dodoi{10.1086/518307}

\bibitem[{{Clayton} {et~al.}(2006){Clayton}, {Kerber}, {Pirzkal}, {De Marco}, {Crowther}, \& {Fedrow}}]{Clayton2006}
{Clayton}, G.~C., {Kerber}, F., {Pirzkal}, N., {et~al.} 2006, \apjl, 646, L69, \dodoi{10.1086/506593}

\bibitem[{{Cox} {et~al.}(1980){Cox}, {Wheeler}, {Hansen}, {King}, {Cox}, \& {Hodson}}]{Cox1980}
{Cox}, J.~P., {Wheeler}, J.~C., {Hansen}, C.~J., {et~al.} 1980, \ssr, 27, 529, \dodoi{10.1007/BF00168346}

\bibitem[{{Crawford} {et~al.}(2020){Crawford}, {Clayton}, {Munson}, {Chatzopoulos}, \& {Frank}}]{Crawford2020}
{Crawford}, C.~L., {Clayton}, G.~C., {Munson}, B., {Chatzopoulos}, E., \& {Frank}, J. 2020, \mnras, 498, 2912, \dodoi{10.1093/mnras/staa2526}

\bibitem[{{Crawford} {et~al.}(2023){Crawford}, {Tisserand}, {Clayton}, {Soon}, {Bessell}, {Wood}, {Garc{\'\i}a-Hern{\'a}ndez}, {Ruiter}, \& {Seitenzahl}}]{Crawford2023}
{Crawford}, C.~L., {Tisserand}, P., {Clayton}, G.~C., {et~al.} 2023, \mnras, 521, 1674, \dodoi{10.1093/mnras/stad324}

\bibitem[{{Cunha} {et~al.}(2015){Cunha}, {Stello}, {Avelino}, {Christensen-Dalsgaard}, \& {Townsend}}]{Cunha2015}
{Cunha}, M.~S., {Stello}, D., {Avelino}, P.~P., {Christensen-Dalsgaard}, J., \& {Townsend}, R.~H.~D. 2015, \apj, 805, 127, \dodoi{10.1088/0004-637X/805/2/127}

\bibitem[{{De} {et~al.}(2020){De}, {Hankins}, {Kasliwal}, {Moore}, {Ofek}, {Adams}, {Ashley}, {Babul}, {Bagdasaryan}, {Burdge}, {Burnham}, {Dekany}, {Declacroix}, {Galla}, {Greffe}, {Hale}, {Jencson}, {Lau}, {Mahabal}, {McKenna}, {Sharma}, {Shopbell}, {Smith}, {Soon}, {Sokoloski}, {Soria}, \& {Travouillon}}]{De2020}
{De}, K., {Hankins}, M.~J., {Kasliwal}, M.~M., {et~al.} 2020, \pasp, 132, 025001, \dodoi{10.1088/1538-3873/ab6069}

\bibitem[{{De Ridder} {et~al.}(2009){De Ridder}, {Barban}, {Baudin}, {Carrier}, {Hatzes}, {Hekker}, {Kallinger}, {Weiss}, {Baglin}, {Auvergne}, {Samadi}, {Barge}, \& {Deleuil}}]{deRidder2009}
{De Ridder}, J., {Barban}, C., {Baudin}, F., {et~al.} 2009, \nat, 459, 398, \dodoi{10.1038/nature08022}

\bibitem[{{Dupret} {et~al.}(2009){Dupret}, {Belkacem}, {Samadi}, {Montalban}, {Moreira}, {Miglio}, {Godart}, {Ventura}, {Ludwig}, {Grigahc{\`e}ne}, {Goupil}, {Noels}, \& {Caffau}}]{Dupret2009}
{Dupret}, M.~A., {Belkacem}, K., {Samadi}, R., {et~al.} 2009, \aap, 506, 57, \dodoi{10.1051/0004-6361/200911713}

\bibitem[{{Feast} {et~al.}(2019){Feast}, {Griffin}, {Herbig}, \& {Whitelock}}]{Feast2019}
{Feast}, M.~W., {Griffin}, R.~F., {Herbig}, G.~H., \& {Whitelock}, P.~A. 2019, \mnras, 482, 4174, \dodoi{10.1093/mnras/sty2893}

\bibitem[{{Fuller} {et~al.}(2015){Fuller}, {Cantiello}, {Stello}, {Garcia}, \& {Bildsten}}]{Fuller2015}
{Fuller}, J., {Cantiello}, M., {Stello}, D., {Garcia}, R.~A., \& {Bildsten}, L. 2015, Science, 350, 423, \dodoi{10.1126/science.aac6933}

\bibitem[{{Garc{\'\i}a-Hern{\'a}ndez} {et~al.}(2010){Garc{\'\i}a-Hern{\'a}ndez}, {Lambert}, {Kameswara Rao}, {Hinkle}, \& {Eriksson}}]{GarciaHernandez2010}
{Garc{\'\i}a-Hern{\'a}ndez}, D.~A., {Lambert}, D.~L., {Kameswara Rao}, N., {Hinkle}, K.~H., \& {Eriksson}, K. 2010, \apj, 714, 144, \dodoi{10.1088/0004-637X/714/1/144}

\bibitem[{{Gautschy} {et~al.}(1990){Gautschy}, {Glatzel}, {Gautschy}, \& {Glatzel}}]{Gautschy1990}
{Gautschy}, A., {Glatzel}, W., {Gautschy}, A., \& {Glatzel}, W. 1990, \mnras, 245, 597, \dodoi{10.1093/mnras/245.4.597}

\bibitem[{{Glatzel} \& {Gautschy}(1992)}]{Glatzel1992}
{Glatzel}, W., \& {Gautschy}, A. 1992, \mnras, 256, 209, \dodoi{10.1093/mnras/256.2.209}

\bibitem[{{Goldreich} \& {Keeley}(1977)}]{Goldreich1977}
{Goldreich}, P., \& {Keeley}, D.~A. 1977, \apj, 212, 243, \dodoi{10.1086/155043}

\bibitem[{{Goldreich} {et~al.}(1994){Goldreich}, {Murray}, \& {Kumar}}]{Goldreich1994}
{Goldreich}, P., {Murray}, N., \& {Kumar}, P. 1994, \apj, 424, 466, \dodoi{10.1086/173904}

\bibitem[{{Goldstein} \& {Townsend}(2020)}]{GYREIII}
{Goldstein}, J., \& {Townsend}, R.~H.~D. 2020, \apj, 899, 116, \dodoi{10.3847/1538-4357/aba748}

\bibitem[{{Harris} {et~al.}(2020){Harris}, {Millman}, {van der Walt}, {Gommers}, {Virtanen}, {Cournapeau}, {Wieser}, {Taylor}, {Berg}, {Smith}, {Kern}, {Picus}, {Hoyer}, {van Kerkwijk}, {Brett}, {Haldane}, {del R{\'\i}o}, {Wiebe}, {Peterson}, {G{\'e}rard-Marchant}, {Sheppard}, {Reddy}, {Weckesser}, {Abbasi}, {Gohlke}, \& {Oliphant}}]{numpy2020}
{Harris}, C.~R., {Millman}, K.~J., {van der Walt}, S.~J., {et~al.} 2020, \nat, 585, 357, \dodoi{10.1038/s41586-020-2649-2}

\bibitem[{{Huber} {et~al.}(2011){Huber}, {Bedding}, {Stello}, {Hekker}, {Mathur}, {Mosser}, {Verner}, {Bonanno}, {Buzasi}, {Campante}, {Elsworth}, {Hale}, {Kallinger}, {Silva Aguirre}, {Chaplin}, {De Ridder}, {Garc{\'\i}a}, {Appourchaux}, {Frandsen}, {Houdek}, {Molenda-{\.Z}akowicz}, {Monteiro}, {Christensen-Dalsgaard}, {Gilliland}, {Kawaler}, {Kjeldsen}, {Broomhall}, {Corsaro}, {Salabert}, {Sanderfer}, {Seader}, \& {Smith}}]{Huber2011}
{Huber}, D., {Bedding}, T.~R., {Stello}, D., {et~al.} 2011, \apj, 743, 143, \dodoi{10.1088/0004-637X/743/2/143}

\bibitem[{Hunter(2007)}]{hunter_2007_aa}
Hunter, J.~D. 2007, Computing In Science \&amp; Engineering, 9, 90

\bibitem[{{Iben} {et~al.}(1996){Iben}, {Tutukov}, \& {Yungelson}}]{Iben1996}
{Iben}, Icko, J., {Tutukov}, A.~V., \& {Yungelson}, L.~R. 1996, \apj, 456, 750, \dodoi{10.1086/176694}

\bibitem[{{Jermyn} {et~al.}(2023){Jermyn}, {Bauer}, {Schwab}, {Farmer}, {Ball}, {Bellinger}, {Dotter}, {Joyce}, {Marchant}, {Mombarg}, {Wolf}, {Sunny Wong}, {Cinquegrana}, {Farrell}, {Smolec}, {Thoul}, {Cantiello}, {Herwig}, {Toloza}, {Bildsten}, {Townsend}, \& {Timmes}}]{MESAVI}
{Jermyn}, A.~S., {Bauer}, E.~B., {Schwab}, J., {et~al.} 2023, \apjs, 265, 15, \dodoi{10.3847/1538-4365/acae8d}

\bibitem[{{Jiang} {et~al.}(2018){Jiang}, {Cantiello}, {Bildsten}, {Quataert}, {Blaes}, \& {Stone}}]{Jiang2018}
{Jiang}, Y.-F., {Cantiello}, M., {Bildsten}, L., {et~al.} 2018, \nat, 561, 498, \dodoi{10.1038/s41586-018-0525-0}

\bibitem[{{Karakas} {et~al.}(2015){Karakas}, {Ruiter}, \& {Hampel}}]{Karakas2015}
{Karakas}, A.~I., {Ruiter}, A.~J., \& {Hampel}, M. 2015, \apj, 809, 184, \dodoi{10.1088/0004-637X/809/2/184}

\bibitem[{{Karambelkar} {et~al.}(2022){Karambelkar}, {Kasliwal}, {Tisserand}, {Clayton}, {Crawford}, {Anand}, {Geballe}, \& {Montiel}}]{Karambelkar2022}
{Karambelkar}, V., {Kasliwal}, M.~M., {Tisserand}, P., {et~al.} 2022, \aap, 667, A84, \dodoi{10.1051/0004-6361/202142918}

\bibitem[{{Karambelkar} {et~al.}(2021){Karambelkar}, {Kasliwal}, {Tisserand}, {De}, {Anand}, {Ashley}, {Delacroix}, {Hankins}, {Jencson}, {Lau}, {McKenna}, {Moore}, {Ofek}, {Smith}, {Soria}, {Soon}, {Tinyanont}, {Travouillon}, \& {Yao}}]{Karambelkar2021}
{Karambelkar}, V.~R., {Kasliwal}, M.~M., {Tisserand}, P., {et~al.} 2021, \apj, 910, 132, \dodoi{10.3847/1538-4357/abe5aa}

\bibitem[{{Kjeldsen} \& {Bedding}(1995)}]{Kjeldsen1995}
{Kjeldsen}, H., \& {Bedding}, T.~R. 1995, \aap, 293, 87, \dodoi{10.48550/arXiv.astro-ph/9403015}

\bibitem[{Kluyver {et~al.}(2016)Kluyver, Ragan-Kelley, P{\'e}rez, Granger, Bussonnier, Frederic, Kelley, Hamrick, Grout, Corlay, {et~al.}}]{kluyver_2016_aa}
Kluyver, T., Ragan-Kelley, B., P{\'e}rez, F., {et~al.} 2016, in Positioning and Power in Academic Publishing: Players, Agents and Agendas: Proceedings of the 20th International Conference on Electronic Publishing, IOS Press, 87

\bibitem[{{Kochanek} {et~al.}(2017){Kochanek}, {Shappee}, {Stanek}, {Holoien}, {Thompson}, {Prieto}, {Dong}, {Shields}, {Will}, {Britt}, {Perzanowski}, \& {Pojma{\'n}ski}}]{Kochanek2017}
{Kochanek}, C.~S., {Shappee}, B.~J., {Stanek}, K.~Z., {et~al.} 2017, \pasp, 129, 104502, \dodoi{10.1088/1538-3873/aa80d9}

\bibitem[{{Lauer} {et~al.}(2019){Lauer}, {Chatzopoulos}, {Clayton}, {Frank}, \& {Marcello}}]{Lauer2019}
{Lauer}, A., {Chatzopoulos}, E., {Clayton}, G.~C., {Frank}, J., \& {Marcello}, D.~C. 2019, \mnras, 488, 438, \dodoi{10.1093/mnras/stz1732}

\bibitem[{{Lawson} \& {Cottrell}(1997)}]{Lawson1997}
{Lawson}, W.~A., \& {Cottrell}, P.~L. 1997, \mnras, 285, 266, \dodoi{10.1093/mnras/285.2.266}

\bibitem[{{Lawson} {et~al.}(1994){Lawson}, {Cottrell}, {Kilkenny}, {Gilmore}, {Kilmartin}, {Marang}, {Roberts}, \& {van Wyk}}]{Lawson1994}
{Lawson}, W.~A., {Cottrell}, P.~L., {Kilkenny}, D., {et~al.} 1994, \mnras, 271, 919, \dodoi{10.1093/mnras/271.4.919}

\bibitem[{{Lawson} {et~al.}(1990){Lawson}, {Cottrelll}, {Kilmartin}, \& {Gilmore}}]{Lawson1990}
{Lawson}, W.~A., {Cottrelll}, P.~L., {Kilmartin}, P.~M., \& {Gilmore}, A.~C. 1990, \mnras, 247, 91

\bibitem[{{Marigo} \& {Aringer}(2009)}]{Marigo2009}
{Marigo}, P., \& {Aringer}, B. 2009, \aap, 508, 1539, \dodoi{10.1051/0004-6361/200912598}

\bibitem[{{Menon} {et~al.}(2013){Menon}, {Herwig}, {Denissenkov}, {Clayton}, {Staff}, {Pignatari}, \& {Paxton}}]{Menon2013}
{Menon}, A., {Herwig}, F., {Denissenkov}, P.~A., {et~al.} 2013, \apj, 772, 59, \dodoi{10.1088/0004-637X/772/1/59}

\bibitem[{{Menon} {et~al.}(2019){Menon}, {Karakas}, {Lugaro}, {Doherty}, \& {Ritter}}]{Menon2019}
{Menon}, A., {Karakas}, A.~I., {Lugaro}, M., {Doherty}, C.~L., \& {Ritter}, C. 2019, \mnras, 482, 2320, \dodoi{10.1093/mnras/sty2606}

\bibitem[{{Mosser} {et~al.}(2012{\natexlab{a}}){Mosser}, {Goupil}, {Belkacem}, {Michel}, {Stello}, {Marques}, {Elsworth}, {Barban}, {Beck}, {Bedding}, {De Ridder}, {Garc{\'\i}a}, {Hekker}, {Kallinger}, {Samadi}, {Stumpe}, {Barclay}, \& {Burke}}]{Mosser2012_core}
{Mosser}, B., {Goupil}, M.~J., {Belkacem}, K., {et~al.} 2012{\natexlab{a}}, \aap, 540, A143, \dodoi{10.1051/0004-6361/201118519}

\bibitem[{{Mosser} {et~al.}(2012{\natexlab{b}}){Mosser}, {Goupil}, {Belkacem}, {Marques}, {Beck}, {Bloemen}, {De Ridder}, {Barban}, {Deheuvels}, {Elsworth}, {Hekker}, {Kallinger}, {Ouazzani}, {Pinsonneault}, {Samadi}, {Stello}, {Garc{\'\i}a}, {Klaus}, {Li}, {Mathur}, \& {Morris}}]{Mosser2012_spin}
---. 2012{\natexlab{b}}, \aap, 548, A10, \dodoi{10.1051/0004-6361/201220106}

\bibitem[{{Mosser} {et~al.}(2013){Mosser}, {Dziembowski}, {Belkacem}, {Goupil}, {Michel}, {Samadi}, {Soszy{\'n}ski}, {Vrard}, {Elsworth}, {Hekker}, \& {Mathur}}]{Mosser2013}
{Mosser}, B., {Dziembowski}, W.~A., {Belkacem}, K., {et~al.} 2013, \aap, 559, A137, \dodoi{10.1051/0004-6361/201322243}

\bibitem[{{Munson} {et~al.}(2022){Munson}, {Chatzopoulos}, \& {Denissenkov}}]{Munson2022}
{Munson}, B., {Chatzopoulos}, E., \& {Denissenkov}, P.~A. 2022, \apj, 939, 45, \dodoi{10.3847/1538-4357/ac9476}

\bibitem[{{Munson} {et~al.}(2021){Munson}, {Chatzopoulos}, {Frank}, {Clayton}, {Crawford}, {Denissenkov}, \& {Herwig}}]{Munson2021}
{Munson}, B., {Chatzopoulos}, E., {Frank}, J., {et~al.} 2021, \apj, 911, 103, \dodoi{10.3847/1538-4357/abeb6c}

\bibitem[{{Osaki}(1975)}]{Osaki1975}
{Osaki}, Y. 1975, \pasj, 27, 237

\bibitem[{{Paxton} {et~al.}(2011){Paxton}, {Bildsten}, {Dotter}, {Herwig}, {Lesaffre}, \& {Timmes}}]{MESAI}
{Paxton}, B., {Bildsten}, L., {Dotter}, A., {et~al.} 2011, \apjs, 192, 3, \dodoi{10.1088/0067-0049/192/1/3}

\bibitem[{{Paxton} {et~al.}(2013){Paxton}, {Cantiello}, {Arras}, {Bildsten}, {Brown}, {Dotter}, {Mankovich}, {Montgomery}, {Stello}, {Timmes}, \& {Townsend}}]{MESAII}
{Paxton}, B., {Cantiello}, M., {Arras}, P., {et~al.} 2013, \apjs, 208, 4, \dodoi{10.1088/0067-0049/208/1/4}

\bibitem[{{Paxton} {et~al.}(2015){Paxton}, {Marchant}, {Schwab}, {Bauer}, {Bildsten}, {Cantiello}, {Dessart}, {Farmer}, {Hu}, {Langer}, {Townsend}, {Townsley}, \& {Timmes}}]{MESAIII}
{Paxton}, B., {Marchant}, P., {Schwab}, J., {et~al.} 2015, \apjs, 220, 15, \dodoi{10.1088/0067-0049/220/1/15}

\bibitem[{{Paxton} {et~al.}(2018){Paxton}, {Schwab}, {Bauer}, {Bildsten}, {Blinnikov}, {Duffell}, {Farmer}, {Goldberg}, {Marchant}, {Sorokina}, {Thoul}, {Townsend}, \& {Timmes}}]{MESAIV}
{Paxton}, B., {Schwab}, J., {Bauer}, E.~B., {et~al.} 2018, \apjs, 234, 34, \dodoi{10.3847/1538-4365/aaa5a8}

\bibitem[{{Paxton} {et~al.}(2019){Paxton}, {Smolec}, {Schwab}, {Gautschy}, {Bildsten}, {Cantiello}, {Dotter}, {Farmer}, {Goldberg}, {Jermyn}, {Kanbur}, {Marchant}, {Thoul}, {Townsend}, {Wolf}, {Zhang}, \& {Timmes}}]{MESAV}
{Paxton}, B., {Smolec}, R., {Schwab}, J., {et~al.} 2019, \apjs, 243, 10, \dodoi{10.3847/1538-4365/ab2241}

\bibitem[{{Percy}(2023)}]{Percy2023}
{Percy}, J.~R. 2023, \jaavso, 51, 64

\bibitem[{{Percy} {et~al.}(2004){Percy}, {Bandara}, {Fernie}, {Cottrell}, \& {Skuljan}}]{Percy2004}
{Percy}, J.~R., {Bandara}, K., {Fernie}, J.~D., {Cottrell}, P.~L., \& {Skuljan}, L. 2004, \jaavso, 33, 27

\bibitem[{P{\'e}rez \& Granger(2007)}]{perez_2007_aa}
P{\'e}rez, F., \& Granger, B.~E. 2007, Computing in Science \& Engineering, 9, 21

\bibitem[{{Rao} \& {Lambert}(2015)}]{Rao2015}
{Rao}, N.~K., \& {Lambert}, D.~L. 2015, \mnras, 447, 3664, \dodoi{10.1093/mnras/stu2748}

\bibitem[{{Rauer} {et~al.}(2014){Rauer}, {Catala}, {Aerts}, {Appourchaux}, {Benz}, {Brandeker}, {Christensen-Dalsgaard}, {Deleuil}, {Gizon}, {Goupil}, {G{\"u}del}, {Janot-Pacheco}, {Mas-Hesse}, {Pagano}, {Piotto}, {Pollacco}, {Santos}, {Smith}, {Su{\'a}rez}, {Szab{\'o}}, {Udry}, {Adibekyan}, {Alibert}, {Almenara}, {Amaro-Seoane}, {Eiff}, {Asplund}, {Antonello}, {Barnes}, {Baudin}, {Belkacem}, {Bergemann}, {Bihain}, {Birch}, {Bonfils}, {Boisse}, {Bonomo}, {Borsa}, {Brand{\~a}o}, {Brocato}, {Brun}, {Burleigh}, {Burston}, {Cabrera}, {Cassisi}, {Chaplin}, {Charpinet}, {Chiappini}, {Church}, {Csizmadia}, {Cunha}, {Damasso}, {Davies}, {Deeg}, {D{\'\i}az}, {Dreizler}, {Dreyer}, {Eggenberger}, {Ehrenreich}, {Eigm{\"u}ller}, {Erikson}, {Farmer}, {Feltzing}, {de Oliveira Fialho}, {Figueira}, {Forveille}, {Fridlund}, {Garc{\'\i}a}, {Giommi}, {Giuffrida}, {Godolt}, {Gomes da Silva}, {Granzer}, {Grenfell}, {Grotsch-Noels}, {G{\"u}nther}, {Haswell}, {Hatzes}, {H{\'e}brard}, {Hekker}, {Helled}, {Heng}, {Jenkins},
  {Johansen}, {Khodachenko}, {Kislyakova}, {Kley}, {Kolb}, {Krivova}, {Kupka}, {Lammer}, {Lanza}, {Lebreton}, {Magrin}, {Marcos-Arenal}, {Marrese}, {Marques}, {Martins}, {Mathis}, {Mathur}, {Messina}, {Miglio}, {Montalban}, {Montalto}, {Monteiro}, {Moradi}, {Moravveji}, {Mordasini}, {Morel}, {Mortier}, {Nascimbeni}, {Nelson}, {Nielsen}, {Noack}, {Norton}, {Ofir}, {Oshagh}, {Ouazzani}, {P{\'a}pics}, {Parro}, {Petit}, {Plez}, {Poretti}, {Quirrenbach}, {Ragazzoni}, {Raimondo}, {Rainer}, {Reese}, {Redmer}, {Reffert}, {Rojas-Ayala}, {Roxburgh}, {Salmon}, {Santerne}, {Schneider}, {Schou}, {Schuh}, {Schunker}, {Silva-Valio}, {Silvotti}, {Skillen}, {Snellen}, {Sohl}, {Sousa}, {Sozzetti}, {Stello}, {Strassmeier}, {{\v{S}}vanda}, {Szab{\'o}}, {Tkachenko}, {Valencia}, {Van Grootel}, {Vauclair}, {Ventura}, {Wagner}, {Walton}, {Weingrill}, {Werner}, {Wheatley}, \& {Zwintz}}]{Rauer2014}
{Rauer}, H., {Catala}, C., {Aerts}, C., {et~al.} 2014, Experimental Astronomy, 38, 249, \dodoi{10.1007/s10686-014-9383-4}

\bibitem[{{Ricker} {et~al.}(2014){Ricker}, {Winn}, {Vanderspek}, {Latham}, {Bakos}, {Bean}, {Berta-Thompson}, {Brown}, {Buchhave}, {Butler}, {Butler}, {Chaplin}, {Charbonneau}, {Christensen-Dalsgaard}, {Clampin}, {Deming}, {Doty}, {De Lee}, {Dressing}, {Dunham}, {Endl}, {Fressin}, {Ge}, {Henning}, {Holman}, {Howard}, {Ida}, {Jenkins}, {Jernigan}, {Johnson}, {Kaltenegger}, {Kawai}, {Kjeldsen}, {Laughlin}, {Levine}, {Lin}, {Lissauer}, {MacQueen}, {Marcy}, {McCullough}, {Morton}, {Narita}, {Paegert}, {Palle}, {Pepe}, {Pepper}, {Quirrenbach}, {Rinehart}, {Sasselov}, {Sato}, {Seager}, {Sozzetti}, {Stassun}, {Sullivan}, {Szentgyorgyi}, {Torres}, {Udry}, \& {Villasenor}}]{Ricker2014}
{Ricker}, G.~R., {Winn}, J.~N., {Vanderspek}, R., {et~al.} 2014, in Society of Photo-Optical Instrumentation Engineers (SPIE) Conference Series, Vol. 9143, Space Telescopes and Instrumentation 2014: Optical, Infrared, and Millimeter Wave, ed. J.~{Oschmann}, Jacobus~M., M.~{Clampin}, G.~G. {Fazio}, \& H.~A. {MacEwen}, 914320, \dodoi{10.1117/12.2063489}

\bibitem[{{Ruiter} {et~al.}(2009){Ruiter}, {Belczynski}, \& {Fryer}}]{Ruiter2009}
{Ruiter}, A.~J., {Belczynski}, K., \& {Fryer}, C. 2009, \apj, 699, 2026, \dodoi{10.1088/0004-637X/699/2/2026}

\bibitem[{{Saio}(2008)}]{Saio2008}
{Saio}, H. 2008, in Astronomical Society of the Pacific Conference Series, Vol. 391, Hydrogen-Deficient Stars, ed. A.~{Werner} \& T.~{Rauch}, 69

\bibitem[{{Saio}(2009)}]{Saio2009}
{Saio}, H. 2009, Communications in Asteroseismology, 158, 245

\bibitem[{{Saio} {et~al.}(1998){Saio}, {Baker}, \& {Gautschy}}]{Saio1998}
{Saio}, H., {Baker}, N.~H., \& {Gautschy}, A. 1998, \mnras, 294, 622, \dodoi{10.1111/j.1365-8711.1998.01195.x}

\bibitem[{{Saio} \& {Jeffery}(2002)}]{Saio2002}
{Saio}, H., \& {Jeffery}, C.~S. 2002, \mnras, 333, 121, \dodoi{10.1046/j.1365-8711.2002.05384.x}

\bibitem[{{Saio} {et~al.}(1984){Saio}, {Wheeler}, \& {Cox}}]{Saio1984}
{Saio}, H., {Wheeler}, J.~C., \& {Cox}, J.~P. 1984, \apj, 281, 318, \dodoi{10.1086/162102}

\bibitem[{{Samadi} \& {Goupil}(2001)}]{Samadi2001}
{Samadi}, R., \& {Goupil}, M.~J. 2001, \aap, 370, 136, \dodoi{10.1051/0004-6361:20010212}

\bibitem[{{Schwab}(2019)}]{Schwab2019}
{Schwab}, J. 2019, \apj, 885, 27, \dodoi{10.3847/1538-4357/ab425d}

\bibitem[{{Shappee} {et~al.}(2014){Shappee}, {Prieto}, {Grupe}, {Kochanek}, {Stanek}, {De Rosa}, {Mathur}, {Zu}, {Peterson}, {Pogge}, {Komossa}, {Im}, {Jencson}, {Holoien}, {Basu}, {Beacom}, {Szczygie{\l}}, {Brimacombe}, {Adams}, {Campillay}, {Choi}, {Contreras}, {Dietrich}, {Dubberley}, {Elphick}, {Foale}, {Giustini}, {Gonzalez}, {Hawkins}, {Howell}, {Hsiao}, {Koss}, {Leighly}, {Morrell}, {Mudd}, {Mullins}, {Nugent}, {Parrent}, {Phillips}, {Pojmanski}, {Rosing}, {Ross}, {Sand}, {Terndrup}, {Valenti}, {Walker}, \& {Yoon}}]{Shappee2014}
{Shappee}, B.~J., {Prieto}, J.~L., {Grupe}, D., {et~al.} 2014, \apj, 788, 48, \dodoi{10.1088/0004-637X/788/1/48}

\bibitem[{{Stello} {et~al.}(2016){Stello}, {Cantiello}, {Fuller}, {Huber}, {Garc{\'\i}a}, {Bedding}, {Bildsten}, \& {Silva Aguirre}}]{Stello2016}
{Stello}, D., {Cantiello}, M., {Fuller}, J., {et~al.} 2016, \nat, 529, 364, \dodoi{10.1038/nature16171}

\bibitem[{{Stello} {et~al.}(2014){Stello}, {Compton}, {Bedding}, {Christensen-Dalsgaard}, {Kiss}, {Kjeldsen}, {Bellamy}, {Garc{\'\i}a}, \& {Mathur}}]{Stello2014}
{Stello}, D., {Compton}, D.~L., {Bedding}, T.~R., {et~al.} 2014, \apjl, 788, L10, \dodoi{10.1088/2041-8205/788/1/L10}

\bibitem[{{Takata}(2006)}]{Takata2006}
{Takata}, M. 2006, \pasj, 58, 893, \dodoi{10.1093/pasj/58.5.893}

\bibitem[{{Tisserand} {et~al.}(2023){Tisserand}, {Crawford}, {Soon}, {Clayton}, {Ruiter}, \& {Seitenzahl}}]{Tisserand2023}
{Tisserand}, P., {Crawford}, C.~L., {Soon}, J., {et~al.} 2023, arXiv e-prints, arXiv:2309.10139, \dodoi{10.48550/arXiv.2309.10139}

\bibitem[{{Tisserand} {et~al.}(2009){Tisserand}, {Wood}, {Marquette}, {Afonso}, {Albert}, {Andersen}, {Ansari}, {Aubourg}, {Bareyre}, {Beaulieu}, {Charlot}, {Coutures}, {Ferlet}, {Fouqu{\'e}}, {Glicenstein}, {Goldman}, {Gould}, {Gros}, {de Kat}, {Lesquoy}, {Loup}, {Magneville}, {Maurice}, {Maury}, {Milsztajn}, {Moniez}, {Palanque-Delabrouille}, {Perdereau}, {Rich}, {Schwemling}, {Spiro}, \& {Vidal-Madjar}}]{Tisserand2009}
{Tisserand}, P., {Wood}, P.~R., {Marquette}, J.~B., {et~al.} 2009, \aap, 501, 985, \dodoi{10.1051/0004-6361/200911808}

\bibitem[{{Tisserand} {et~al.}(2020){Tisserand}, {Clayton}, {Bessell}, {Welch}, {Kamath}, {Wood}, {Wils}, {Wyrzykowski}, {Mr{\'o}z}, \& {Udalski}}]{Tisserand2020}
{Tisserand}, P., {Clayton}, G.~C., {Bessell}, M.~S., {et~al.} 2020, \aap, 635, A14, \dodoi{10.1051/0004-6361/201834410}

\bibitem[{{Tisserand} {et~al.}(2022){Tisserand}, {Crawford}, {Clayton}, {Ruiter}, {Karambelkar}, {Bessell}, {Seitenzahl}, {Kasliwal}, {Soon}, \& {Travouillon}}]{Tisserand2022}
{Tisserand}, P., {Crawford}, C.~L., {Clayton}, G.~C., {et~al.} 2022, \aap, 667, A83, \dodoi{10.1051/0004-6361/202142916}

\bibitem[{{Townsend} {et~al.}(2018){Townsend}, {Goldstein}, \& {Zweibel}}]{GYREII}
{Townsend}, R.~H.~D., {Goldstein}, J., \& {Zweibel}, E.~G. 2018, \mnras, 475, 879, \dodoi{10.1093/mnras/stx3142}

\bibitem[{{Townsend} \& {Teitler}(2013)}]{GYREI}
{Townsend}, R.~H.~D., \& {Teitler}, S.~A. 2013, \mnras, 435, 3406, \dodoi{10.1093/mnras/stt1533}

\bibitem[{{Viani} {et~al.}(2017){Viani}, {Basu}, {Chaplin}, {Davies}, \& {Elsworth}}]{Viani2017}
{Viani}, L.~S., {Basu}, S., {Chaplin}, W.~J., {Davies}, G.~R., \& {Elsworth}, Y. 2017, \apj, 843, 11, \dodoi{10.3847/1538-4357/aa729c}

\bibitem[{{Virtanen} {et~al.}(2020){Virtanen}, {Gommers}, {Oliphant}, {Haberland}, {Reddy}, {Cournapeau}, {Burovski}, {Peterson}, {Weckesser}, {Bright}, {van der Walt}, {Brett}, {Wilson}, {Millman}, {Mayorov}, {Nelson}, {Jones}, {Kern}, {Larson}, {Carey}, {Polat}, {Feng}, {Moore}, {VanderPlas}, {Laxalde}, {Perktold}, {Cimrman}, {Henriksen}, {Quintero}, {Harris}, {Archibald}, {Ribeiro}, {Pedregosa}, {van Mulbregt}, \& {SciPy 1. 0 Contributors}}]{scipy2020}
{Virtanen}, P., {Gommers}, R., {Oliphant}, T.~E., {et~al.} 2020, Nature Methods, 17, 261, \dodoi{10.1038/s41592-019-0686-2}

\bibitem[{{Warner}(1967)}]{Warner1967}
{Warner}, B. 1967, \mnras, 137, 119, \dodoi{10.1093/mnras/137.2.119}

\bibitem[{{Webbink}(1984)}]{Webbink1984}
{Webbink}, R.~F. 1984, \apj, 277, 355, \dodoi{10.1086/161701}

\bibitem[{{Weiss}(1987)}]{Weiss1987_pulsation}
{Weiss}, A. 1987, \aap, 185, 178

\bibitem[{Wolf \& Schwab(2017)}]{bill_wolf_2017_826958}
Wolf, B., \& Schwab, J. 2017, wmwolf/py\_mesa\_reader: Interact with MESA Output, 0.3.0,  Zenodo, \dodoi{10.5281/zenodo.826958}

\bibitem[{{Wood}(1976)}]{Wood1967}
{Wood}, P.~R. 1976, \mnras, 174, 531, \dodoi{10.1093/mnras/174.3.531}

\bibitem[{{Y{\i}ld{\i}z} {et~al.}(2016){Y{\i}ld{\i}z}, {{\c{C}}elik Orhan}, \& {Kayhan}}]{Yildiz2016}
{Y{\i}ld{\i}z}, M., {{\c{C}}elik Orhan}, Z., \& {Kayhan}, C. 2016, \mnras, 462, 1577, \dodoi{10.1093/mnras/stw1709}

\bibitem[{{Yu} {et~al.}(2020){Yu}, {Bedding}, {Stello}, {Huber}, {Compton}, {Gizon}, \& {Hekker}}]{Yu2020}
{Yu}, J., {Bedding}, T.~R., {Stello}, D., {et~al.} 2020, \mnras, 493, 1388, \dodoi{10.1093/mnras/staa300}

\bibitem[{{Yu} {et~al.}(2018){Yu}, {Huber}, {Bedding}, {Stello}, {Hon}, {Murphy}, \& {Khanna}}]{Yu2018}
{Yu}, J., {Huber}, D., {Bedding}, T.~R., {et~al.} 2018, \apjs, 236, 42, \dodoi{10.3847/1538-4365/aaaf74}

\bibitem[{{Zinn} {et~al.}(2023){Zinn}, {Pinsonneault}, {Bildsten}, \& {Stello}}]{Zinn2023}
{Zinn}, J.~C., {Pinsonneault}, M.~H., {Bildsten}, L., \& {Stello}, D. 2023, arXiv e-prints, arXiv:2308.09854.
\newblock \doarXiv{2308.09854}

\end{thebibliography}
\bibliographystyle{aasjournal}

\end{document}